\documentclass[12pt,a4paper]{article}
\usepackage[utf8]{inputenc}
\usepackage[T1]{fontenc}
\usepackage{amsmath,amssymb}
\usepackage{graphicx}
\usepackage{geometry}
\usepackage{cite}
\usepackage{hyperref}
\usepackage{authblk}
\usepackage{microtype}

\hypersetup{colorlinks=true,linkcolor=blue,urlcolor=cyan,citecolor=blue}

\title{\textbf{Modulated structures in discrete approximations to a three-dimensional cholesteric model on a recursive lattice}}
\author[1]{William de Castilho\thanks{Electronic address: william2.castilho@alumni.usp.br}}
\affil[1]{Independent Author, S\~ao Paulo, SP, Brazil}
\date{\today}

\begin{document}
\maketitle

\begin{abstract}
We study finite-state approximations to a chiral nematic lattice model whose local directors are embedded in three-dimensional orientational space. The model actually analyzed is formulated in terms of a rotated quadrupolar interaction, which provides a direct angular chirality parameter $\Delta$ and reduces to the usual planar chiral-nematic form for directors confined to the $xy$ plane. On a rooted Cayley tree the statistical problem is written as a nonlinear recursion for branch partition functions; in the infinite-coordination limit, with $rJ_Q$ fixed, the recursion becomes a softmax map for the state probabilities. We analyze Cartesian three-state, planar four-state, and non-planar four-state discretizations. The first supports ordered and period-two attractors, whereas the planar four-state discretization displays a sequence of commensurate and longer-period modulated attractors. Its disordered fixed point loses stability at $t=9/16$ through a complex-conjugate pair with phase $q=\pm2\Delta$, so a modulated critical mode is selected for every nonzero chirality. The non-planar four-state set is intrinsically anisotropic and, when the stated model is reconstructed directly, supports prominent period-three as well as period-two and higher-period cycles. Because attractor stability does not by itself establish global thermodynamic stability, the diagrams reported here are explicitly interpreted as attractor/stability diagrams rather than equilibrium phase diagrams.

\vspace{0.4cm}
\end{abstract}

\section{Introduction}

Cholesteric, or chiral nematic, liquid crystals possess orientational order without the long-range positional order characteristic of layered liquid-crystalline phases \cite{Genne1993}. Their defining feature is a helical modulation of the nematic director along a distinguished axis, which produces a finite pitch. Closely related competition occurs in magnetic systems with Dzyaloshinskii--Moriya (DM) interactions, where an ordinary alignment tendency competes with a chiral twist \cite{William2021,Izyumov1984}. This analogy motivates minimal lattice models of spatially modulated orientational order.

A commonly used planar pair interaction can be written as
\begin{equation}
\mathcal H_{c,ij}
=-J_N(\vec n_i\!\cdot\!\vec n_j)^2
-D(\vec n_i\!\cdot\!\vec n_j)
 (\vec n_i\!\times\!\vec n_j)\cdot\hat z,
\label{eq:motivating_hamiltonian}
\end{equation}
where $J_N>0$ is a nematic coupling and $D$ gauges chirality. This form is useful as physical motivation, but the transfer matrices used below correspond more directly to a rotated quadrupolar interaction. Making this distinction explicit removes an ambiguity in the parameterization of the model.

The present work extends the recursive-lattice strategy used for planar chiral nematic clock models \cite{Nascimento2019}. We use finite sets of director axes embedded in three dimensions and ask how the orientational discretization affects the attractors of the recursion. The calculation is formulated on a rooted Cayley tree. Deep-interior fixed points and cycles are the cavity solutions usually associated with Bethe-lattice calculations, but the distinction between a finite Cayley tree and an infinite Bethe lattice is important \cite{Ostilli2012}. In particular, the existence of a stable attractor is a statement about the nonlinear recursion; it is not automatically a proof that the corresponding solution minimizes a bulk thermodynamic free energy. We therefore use the terminology \emph{attractor/stability diagram} throughout.

\section{Model and parameterization}

\subsection{Rotated quadrupolar interaction}

The local director is a unit vector $\vec n_i$ with the nematic identification $\vec n_i\equiv-\vec n_i$. We define
\begin{equation}
Q_i=\frac12\left(3\vec n_i\vec n_i^{\,T}-I\right),
\label{eq:Qdef}
\end{equation}
and a rotation about the $z$ axis,
\begin{equation}
R_3(\Delta)=
\begin{pmatrix}
\cos\Delta&\sin\Delta&0\\
-\sin\Delta&\cos\Delta&0\\
0&0&1
\end{pmatrix}.
\label{eq:Rdef}
\end{equation}
The interaction used to construct the transfer matrices is
\begin{equation}
\mathcal H_Q
=-J_Q\sum_{\langle i,j\rangle}
Q_i:\left[R_3(\Delta)Q_jR_3^T(\Delta)\right],
\label{eq:HQ}
\end{equation}
where $A:B\equiv\mathrm{Tr}(A^TB)$. For chiral bonds the ordered pair $(i,j)$ follows the positive axial direction; on the rooted tree this direction is identified with successive generations.

The notation ``three-dimensional'' refers to the orientational space of the director. Because the interaction singles out $\hat z$, Eq.~\eqref{eq:HQ} is not globally $O(3)$ invariant.

\subsection{Connection between $D$, $\Delta$, and $p$}

For planar directors, $\vec n=(\cos\theta,\sin\theta,0)$, let $\delta=\theta_i-\theta_j$. Equation~\eqref{eq:Qdef} gives
\begin{equation}
Q_i:\left[R_3(\Delta)Q_jR_3^T(\Delta)\right]
=\frac38+\frac98\cos\left[2(\delta+\Delta)\right].
\label{eq:Qplanar}
\end{equation}
On the other hand, the pair contribution of Eq.~\eqref{eq:motivating_hamiltonian} is
\begin{equation}
\mathcal H_{c,ij}
=-\frac{J_N}{2}
-\frac{J_N}{2}\cos 2\delta
+\frac{D}{2}\sin 2\delta.
\label{eq:DMplanar}
\end{equation}
Up to an additive constant, Eqs.~\eqref{eq:HQ} and \eqref{eq:DMplanar} agree in the planar subspace if
\begin{equation}
J_N=\frac{9J_Q}{4}\cos 2\Delta,
\qquad
D=\frac{9J_Q}{4}\sin 2\Delta.
\label{eq:parameterization}
\end{equation}
Hence
\begin{equation}
p\equiv\frac{D}{J_N}=\tan 2\Delta,
\qquad
J_Q=\frac49\sqrt{J_N^2+D^2}.
\label{eq:p_relation}
\end{equation}
This mapping is exact only for planar directors. For the non-planar states studied below, $\Delta$ is therefore taken as the primary chirality parameter of Eq.~\eqref{eq:HQ}; $p=D/J_N$ should not be used as though it were an independent exact parameterization of the full three-dimensional model. Scanning the full interval $0\leq\Delta\leq\pi$ likewise explores the periodic rotated-tensor model beyond the sector with simultaneously positive $J_N$ and $D$ in Eq.~\eqref{eq:motivating_hamiltonian}.

\section{Recursive-lattice formulation}

Let the allowed director axes be indexed by $a=1,\ldots,m$, and define
\begin{equation}
V_{ab}(\Delta)
=Q_a:\left[R_3(\Delta)Q_bR_3^T(\Delta)\right],
\qquad
T_{ab}=\exp\left(\beta J_Q V_{ab}\right).
\label{eq:VandT}
\end{equation}
For a rooted Cayley tree of ramification $r$, each non-root site has one parent and $r$ descendants, so the bulk coordination is
\begin{equation}
z=r+1.
\label{eq:coordination}
\end{equation}
If $Z_a^{(n)}$ is the partial partition function of an $n$-generation branch whose root is in state $a$, then
\begin{equation}
Z_a^{(n+1)}
=\left[\sum_{b=1}^{m}T_{ab}Z_b^{(n)}\right]^r.
\label{eq:Zrecursion}
\end{equation}
Introducing normalized branch weights
\begin{equation}
\rho_a^{(n)}=\frac{Z_a^{(n)}}{\sum_b Z_b^{(n)}},
\end{equation}
we obtain the finite-$r$ map
\begin{equation}
\rho'_a
=\frac{\left(\sum_bT_{ab}\rho_b\right)^r}
{\sum_c\left(\sum_bT_{cb}\rho_b\right)^r}.
\label{eq:rhomap_finite}
\end{equation}

Following the standard infinite-coordination construction used in related chiral recursive-lattice calculations \cite{Nascimento2019,William2021}, we take
\begin{equation}
r\rightarrow\infty,
\qquad
J_Q\rightarrow0,
\qquad
\bar J\equiv rJ_Q\quad\text{fixed},
\label{eq:mf_limit}
\end{equation}
and define the reduced temperature
\begin{equation}
t\equiv\frac{1}{\beta\bar J}
=\frac{k_BT_{\rm phys}}{rJ_Q}.
\label{eq:reducedT}
\end{equation}
Expanding Eq.~\eqref{eq:rhomap_finite} to leading order in $1/r$ gives the limiting map
\begin{equation}
\boxed{
\rho'_a
=\frac{\exp\left[(V\rho)_a/t\right]}
{\sum_c\exp\left[(V\rho)_c/t\right]}}
\label{eq:mf_map}
\end{equation}
which is the map used for all numerical diagrams below.

\subsection{Numerical protocol and classification}

The diagrams were regenerated directly from Eq.~\eqref{eq:mf_map}. We used $\Delta$ from $0^\circ$ to $180^\circ$ in steps of $1^\circ$. The reduced-temperature spacing was $\Delta t=0.005$; the three-state diagram used $0.02\leq t\leq0.90$, and the four-state diagrams used $0.02\leq t\leq0.70$ (the non-planar figure is displayed up to $t=0.65$). At each point we started from a deterministic near-symmetric probability vector with a perturbation of order $10^{-4}$, discarded $5000$ iterations, and analyzed the following $256$ iterates.

The smallest period $P\leq64$ was accepted when three consecutive repeated blocks agreed in the maximum norm to $10^{-9}$. Periods $P=2,3,4$ are labeled $1/2$, $1/3$, and $1/4$, respectively; other detected periods and unresolved long-period or slowly converging points are grouped into $M$. A fixed point is labeled ordered (O) when it breaks the symmetry of the corresponding high-temperature branch. For the first two discretizations the disordered state is the uniform distribution. The non-planar four-state set is not isotropic, so its high-temperature symmetry-preserving fixed point is denoted H rather than D.

We also repeated representative points from state-biased and deterministic pseudorandom initial conditions. Multiple attractor classes occur for some parameters. Consequently, basin dependence is real and the recurrence alone does not define thermodynamic coexistence. No Bethe free-energy selection among competing cycles is imposed in this work; terms such as ``coexistence'' and ``first-order transition'' are therefore not inferred solely from multistability.

OpenAI GPT-5.6 Sol (``GPT Sol'') was used as an AI-assisted research tool during the reconstruction and checking of the recurrence-map calculation, including algebraic consistency checks and assistance in developing the reproducible numerical implementation. The resulting equations, code, numerical classifications, and scientific interpretation remain subject to author review and responsibility.

\section{Three-state Cartesian discretization}

The allowed director axes are
\begin{equation}
\vec n_1=\pm(1,0,0),\qquad
\vec n_2=\pm(0,1,0),\qquad
\vec n_3=\pm(0,0,1).
\end{equation}
Equation~\eqref{eq:VandT} gives
\begin{equation}
T_{3}=
\begin{pmatrix}
 e^{\frac38\beta J_Q(1+3\cos2\Delta)} &
 e^{\frac38\beta J_Q(1-3\cos2\Delta)} & e^{-3\beta J_Q/4}\\
 e^{\frac38\beta J_Q(1-3\cos2\Delta)} &
 e^{\frac38\beta J_Q(1+3\cos2\Delta)} & e^{-3\beta J_Q/4}\\
 e^{-3\beta J_Q/4}&e^{-3\beta J_Q/4}&e^{3\beta J_Q/2}
\end{pmatrix},
\label{eq:T3}
\end{equation}
which reproduces the transfer matrix used in the original calculation.

The interaction matrix has eigenvalues
\begin{equation}
0,\qquad \frac94,\qquad \frac94\cos2\Delta.
\end{equation}
Linearizing Eq.~\eqref{eq:mf_map} around the uniform fixed point $\rho_a=1/3$ therefore gives the two nontrivial multipliers
\begin{equation}
\lambda_1=\frac{3}{4t},
\qquad
\lambda_2=\frac{3\cos2\Delta}{4t}.
\label{eq:eig3}
\end{equation}
The high-temperature uniform fixed point is linearly stable for
\begin{equation}
t>\frac34.
\label{eq:t3crit}
\end{equation}
The instability is governed by real multipliers, consistent with the restricted modulation structure of this coarse discretization.

\begin{figure}[htbp]
\centering
\includegraphics[width=0.86\textwidth]{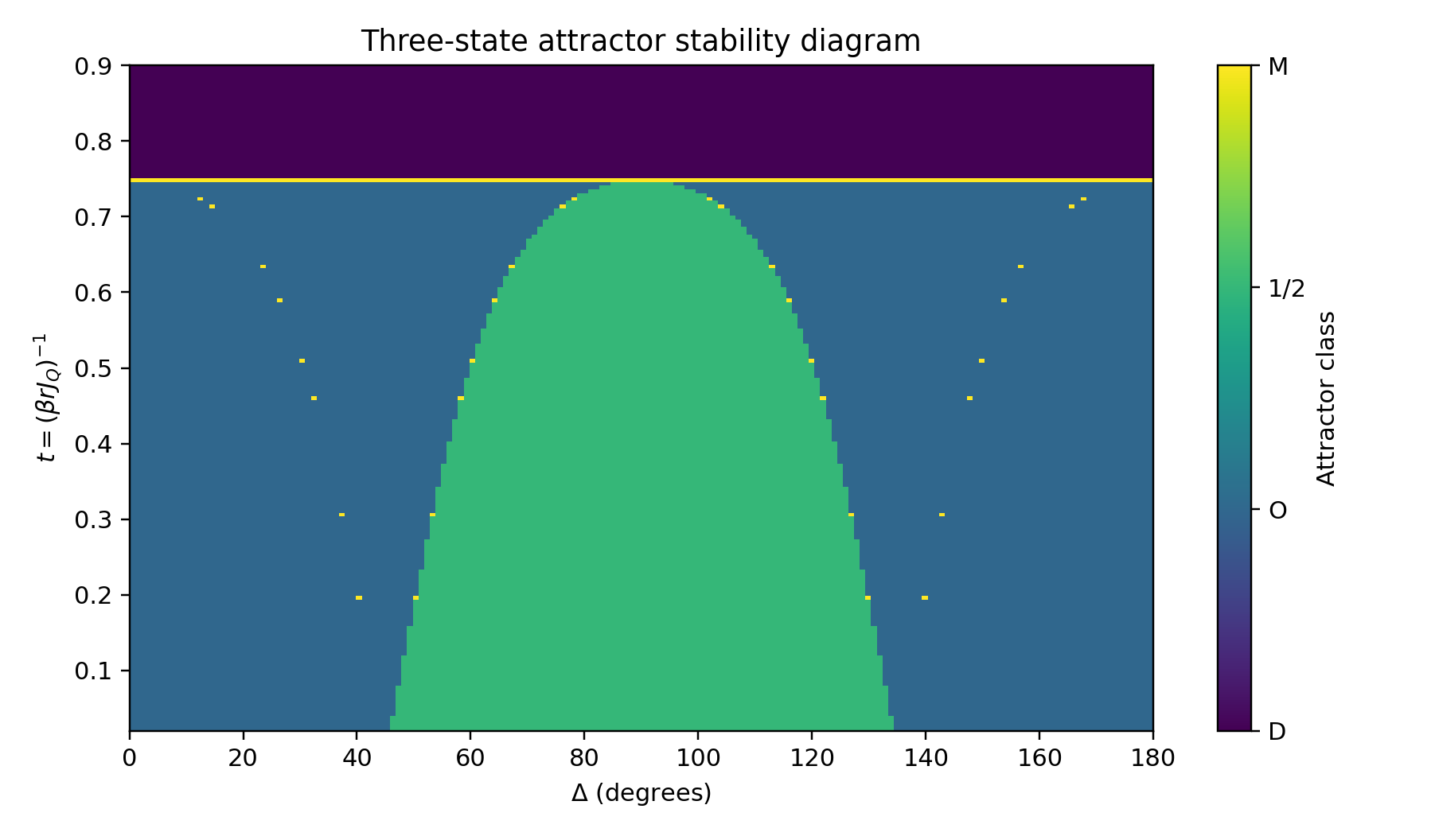}
\caption{Three-state attractor/stability diagram obtained from Eq.~\eqref{eq:mf_map} using the numerical protocol described in the text. D denotes the uniform high-temperature fixed point, O an ordered fixed point, $1/2$ a period-two cycle, and M unresolved or other modulated behavior near stability boundaries. The analytic linear-stability threshold of D is $t=3/4$.}
\label{fig:three}
\end{figure}

Figure~\ref{fig:three} shows the attractor reached from the near-symmetric initial condition. Ordered fixed points occur on the sides of the diagram, while a period-two attractor occupies the central low-temperature region. Multi-start checks show that additional stable attractors can coexist dynamically with these solutions; for example, at $\Delta=90^\circ$ and $t=0.40$, different initial conditions converge either to the period-two cycle or to an ordered fixed point. This is why Fig.~\ref{fig:three} is not interpreted as an equilibrium phase diagram.

\section{Planar four-state discretization}

The planar director axes are
\begin{equation}
\theta_a=(a-1)\frac{\pi}{4},\qquad
\vec n_a=(\cos\theta_a,\sin\theta_a,0),
\qquad a=1,\ldots,4.
\end{equation}
For these states,
\begin{equation}
V_{ab}=\frac38+\frac98
\cos\left[2(\theta_a-\theta_b+\Delta)\right].
\label{eq:V4planar}
\end{equation}
Together with Eqs.~\eqref{eq:VandT}--\eqref{eq:mf_map}, Eq.~\eqref{eq:V4planar} completely specifies both the finite-$r$ recursion and its infinite-coordination limit; no additional unpublished recursion relations are required for reproducibility.

The uniform state $\rho_a=1/4$ is a fixed point. In the three-dimensional probability simplex the linearized map has a zero mode and the complex-conjugate pair
\begin{equation}
\lambda_{\pm}
=\frac{9}{16t}\,e^{\pm i2\Delta}.
\label{eq:lambda4planar}
\end{equation}
Consequently the disordered fixed point is linearly stable for
\begin{equation}
t>\frac{9}{16},
\label{eq:t4crit}
\end{equation}
and its critical oscillatory mode has generation-space wave number
\begin{equation}
q=\arg\lambda_{\pm}=\pm2\Delta\pmod{2\pi}.
\label{eq:qDelta}
\end{equation}
This relation gives a model-specific version of the usual DM result: combining Eq.~\eqref{eq:p_relation} with Eq.~\eqref{eq:qDelta} in the planar sector gives $\tan q=\pm p$, up to the sign convention chosen for the oriented bond.

\begin{figure}[htbp]
\centering
\includegraphics[width=0.90\textwidth]{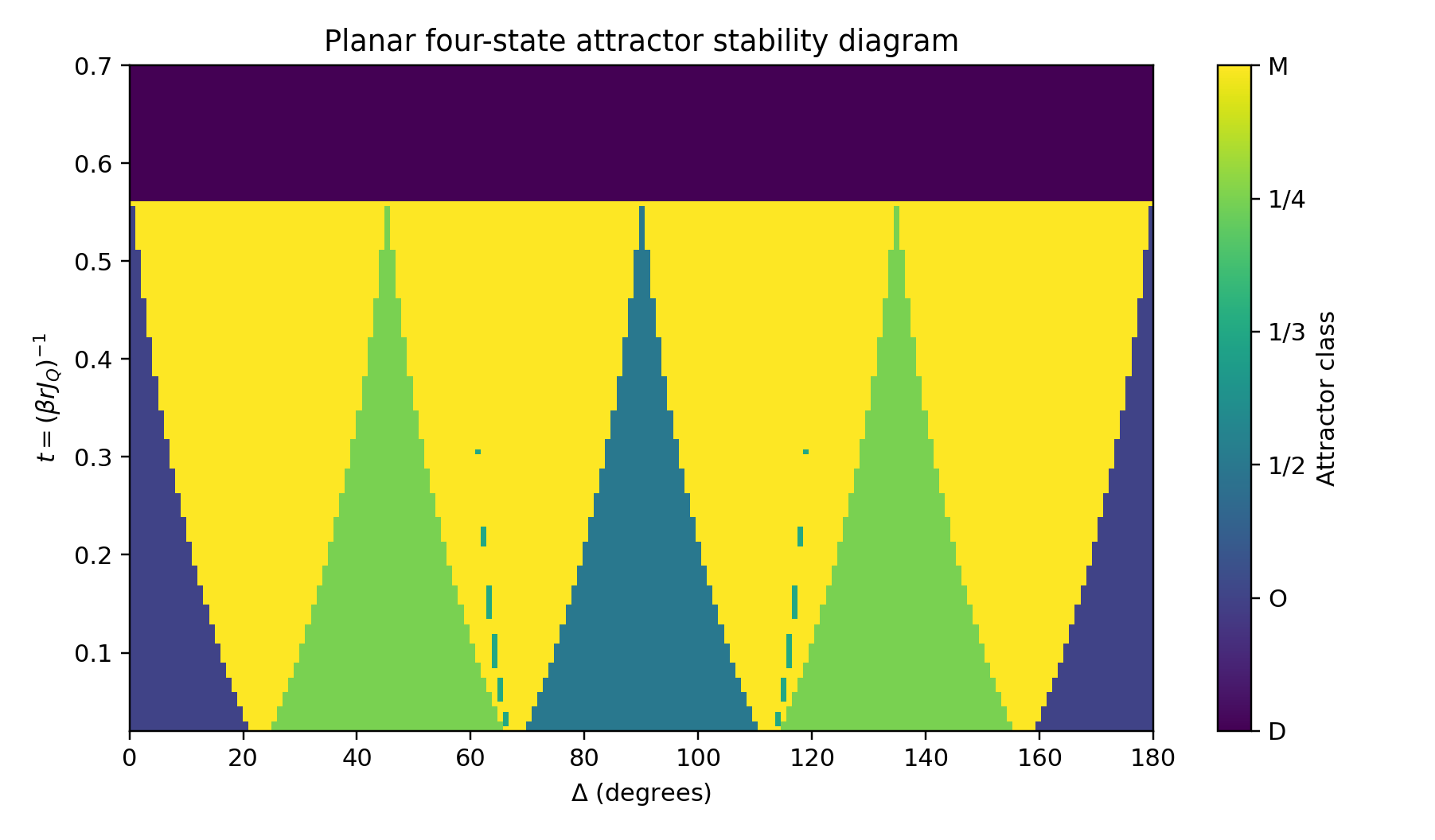}
\caption{Planar four-state attractor/stability diagram reconstructed from Eqs.~\eqref{eq:mf_map} and \eqref{eq:V4planar}. D denotes the uniform fixed point, O an ordered fixed point, $1/2$, $1/3$, and $1/4$ the indicated periods when present, and M all other detected or unresolved modulated behavior. The analytic stability threshold of D is $t=9/16$.}
\label{fig:planar4}
\end{figure}

Figure~\ref{fig:planar4} displays the resulting stability diagram. The commensurate structures follow directly from Eq.~\eqref{eq:qDelta}: for example, $\Delta=45^\circ$ gives $q=\pi/2$ and a period-four cycle, while $\Delta=90^\circ$ gives $q=\pi$ and a period-two cycle. Intermediate angles generate longer-period lock-in and unresolved/long-period regions collected under M. There is no finite-chirality ANNNI-type Lifshitz point in this linear-stability problem: a modulated critical mode is selected for every $\Delta\neq0$ (modulo the discrete symmetries), and the homogeneous $q=0$ limit is reached only at the achiral point.

\section{Non-planar four-state discretization}

The third discretization uses
\begin{equation}
\vec n_1=\pm(1,0,0),\quad
\vec n_2=\pm\left(\frac12,\frac{\sqrt3}{2},0\right),\quad
\vec n_3=\pm\left(-\frac12,\frac{\sqrt3}{2},0\right),\quad
\vec n_4=\pm(0,0,1).
\label{eq:states4np}
\end{equation}
This set is not a regular tetrahedral discretization: three axes are planar and one is axial. Its discrete orientational measure is also not isotropic, since
\begin{equation}
\frac14\sum_{a=1}^{4}\vec n_a\vec n_a^{\,T}
=\mathrm{diag}\left(\frac38,\frac38,\frac14\right)
\neq\frac13 I.
\label{eq:anisotropy4np}
\end{equation}
Thus even the equal-weight distribution carries a residual quadrupolar bias, and the finite-temperature symmetry-preserving fixed point is not exactly the isotropic state. We denote this high-temperature branch H.

\begin{figure}[htbp]
\centering
\includegraphics[width=0.90\textwidth]{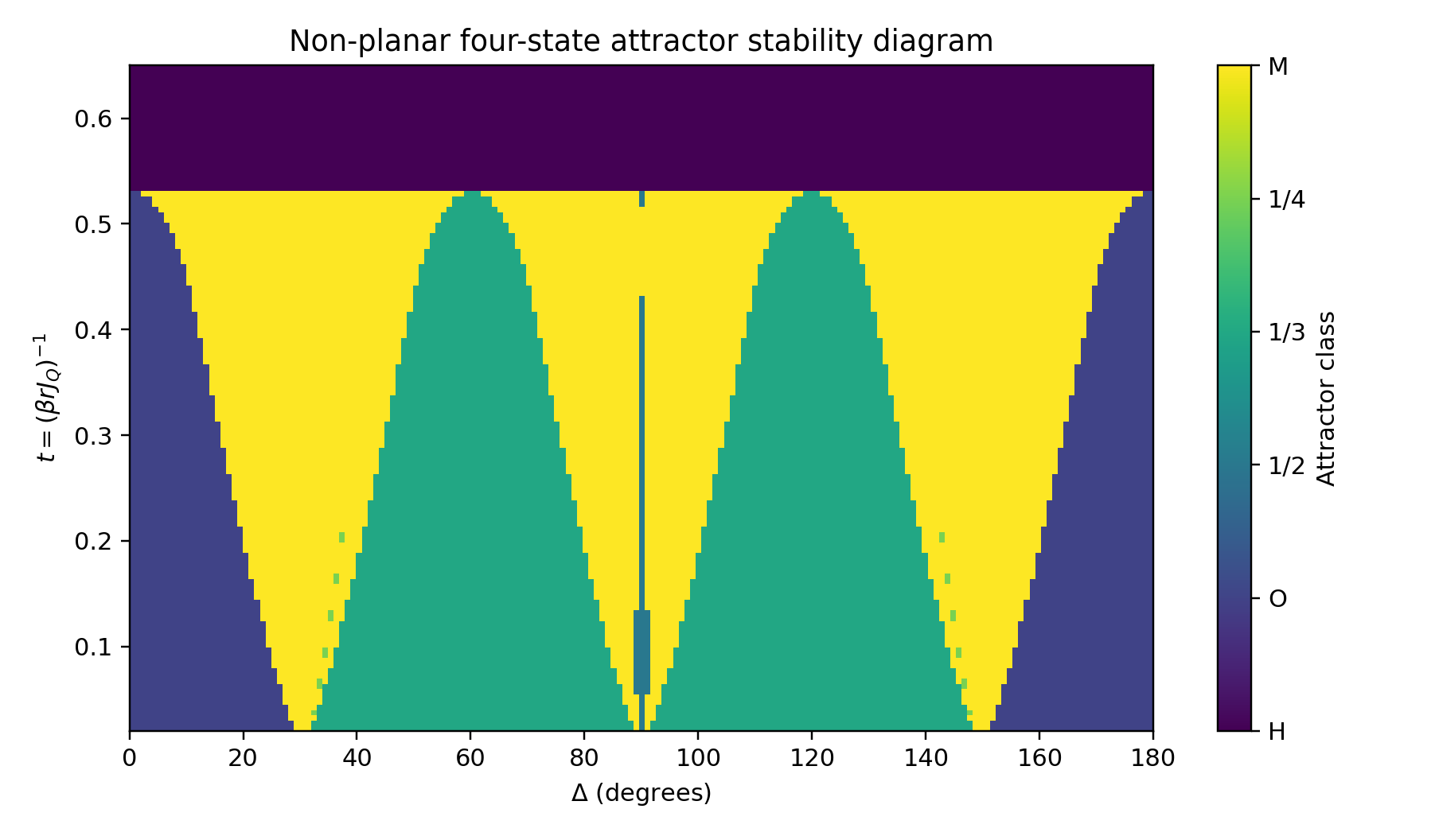}
\caption{Non-planar four-state attractor/stability diagram reconstructed directly from Eqs.~\eqref{eq:mf_map} and \eqref{eq:states4np}. H is the high-temperature symmetry-preserving branch of the anisotropic state set, O an ordered fixed point, and the remaining labels denote periodic or more general modulated attractors. In particular, the stated model produces prominent period-three sectors as well as the period-two structure near $\Delta=90^\circ$.}
\label{fig:nonplanar4}
\end{figure}

The recursion is nevertheless completely determined by Eq.~\eqref{eq:mf_map} with the state set in Eq.~\eqref{eq:states4np}. Reconstructing the map in this way produces the stability diagram in Fig.~\ref{fig:nonplanar4}. A period-two tongue appears around $\Delta=90^\circ$, but broad period-three sectors occur around $\Delta\simeq60^\circ$ and $120^\circ$, together with longer-period structures. This result is a direct consequence of the stated Hamiltonian and state set and should be used as a consistency check on any independently generated diagram for this discretization.

\section{Thermodynamic interpretation}

The recursive map provides a controlled way to locate fixed points, cycles, their basins, and their linear stability. It does not, by itself, assign a global equilibrium free energy to each radial cycle. This distinction is especially important in the present setting because the chiral bond is oriented along successive generations and because different initial conditions can converge to different attractors at the same $(t,\Delta)$.

For this reason, the present results establish \emph{stability regions of recursive solutions}. We do not identify basin multiplicity with thermodynamic coexistence, and we do not infer first-order transition lines solely from the simultaneous existence of attractors. An equilibrium selection would require an additional, explicitly defined bulk free-energy construction that is consistent with the oriented recursive geometry and with periodic generation-dependent solutions. Until such a calculation is supplied, the more conservative stability-diagram terminology is required.

\section{Conclusions}

We have reformulated the discrete three-dimensional cholesteric calculation so that the model parameters, recursive-lattice limit, and numerical classification are explicit and reproducible. The transfer matrices correspond to a rotated quadrupolar interaction, Eq.~\eqref{eq:HQ}. In the planar subspace this model is related to a nematic-plus-chiral pair interaction through $p=D/J_N=\tan2\Delta$, but $\Delta$ remains the primary parameter for non-planar directors.

The tree ramification is $r=z-1$, and the infinite-coordination limit must be taken with $rJ_Q$ fixed. The corresponding reduced temperature is $t=(\beta rJ_Q)^{-1}$. In this limit all three discretizations are generated by the common map in Eq.~\eqref{eq:mf_map}.

The three-state discretization has a disordered stability threshold $t=3/4$ and supports a large period-two attractor region for suitable initial conditions. The planar four-state discretization has a disordered threshold $t=9/16$ and a complex critical pair with $q=\pm2\Delta$, which explains the appearance of period-four, period-two, and longer-period modulated structures. In this planar model the modulated instability exists for every nonzero chirality, so no finite-chirality Lifshitz point of the standard ANNNI type appears in the linear-stability diagram.

The non-planar four-state set is intrinsically anisotropic and should not be described as a regular tetrahedral approximation. Direct reconstruction of the stated model yields prominent period-three and higher-period attractors in addition to the period-two region. This sensitivity to the orientational grid is itself an important result: conclusions about the continuous three-dimensional director problem require a systematic sequence of increasingly isotropic state sets rather than a single low-order discretization.

Finally, all diagrams in this work are stability diagrams of the nonlinear recursion. A future equilibrium analysis would require a separate free-energy prescription capable of comparing competing periodic recursive solutions.

\section*{Acknowledgments}
We acknowledge financial support from the Brazilian agencies CNPq and CAPES.

\section*{Declaration of generative AI and AI-assisted technologies in the research and manuscript preparation process}
During the preparation of this work, the author used OpenAI GPT-5.6 Sol (``GPT Sol'') to assist with analytical consistency checks, reconstruction and verification of the recursive-map numerical calculations, preparation of reproducible numerical code, and revision of the manuscript. After using this tool, the author reviewed and edited the AI-assisted content as needed and takes full responsibility for the content of the publication.

\end{document}